\documentclass[twocolumn,showpacs,amsfonts,aps,prc,floatfix,superscriptaddress,nofootinbib,groupedaddress]{revtex4-1}

\usepackage{graphicx}  
\usepackage{dcolumn}   
\usepackage{bm}        
\usepackage{amssymb}   
\usepackage{wasysym}
\usepackage{hyperref}
\hypersetup{colorlinks=true}
\usepackage{graphicx}
\usepackage{epstopdf}
\usepackage{graphicx}
\usepackage{dcolumn}
\usepackage{bm}
\usepackage{color}
\usepackage{enumerate}

 \newif\ifpdf
 \ifx\pdfoutput\undefined
 \pdffalse
 \else
 \pdfoutput=1
 \pdftrue
 \fi
 \ifpdf
 \usepackage{graphicx}
 \usepackage{epstopdf}
 \else
 \usepackage{graphicx}
 \fi

\newcommand{\pt}{p_{\rm T}}

\newcommand{\gmom}{{\rm GeV}/c}
\newcommand{\sNN}{\sqrt{s_{\rm NN}}}

\begin{document}

\title{Spectra and elliptic flow of (multi-)strange hadrons at RHIC and LHC within viscous hydrodynamics+hadron cascade hybrid model}
\author{Xiangrong Zhu}
\email[Correspond to\ ]{xrongzhu@pku.edu.cn; xrongzhu@zjhu.edu.cn}
\affiliation{School of Science, Huzhou University, Huzhou 313000, P.R. China}
\affiliation{Department of Physics and State Key Laboratory of Nuclear Physics and Technology, Peking
University, Beijing 100871, P.R. China }

\begin{abstract}
Using the (2+1)-dimensional ultrarelativistic viscous hydrodynamics+hadron cascade, {\tt VISHNU}, hybrid model, we study the $\pt$-spectra and elliptic flow of $\Lambda$, $\Xi$, and $\Omega$ in Au+Au collisions at $\sNN$=200 GeV and in 
Pb+Pb collisions at $\sNN$=2.76 TeV. Comparing our model results with the data measurements, we find that the {\tt VISHNU} model gives good descriptions of the measurements of these strange and multi-strange hadrons at several centrality classes at RHIC and LHC. Mass ordering of elliptic flow $v_{2}$ among $\pi$, $K$, $p$, $\Lambda$, $\Xi$, and $\Omega$ are further investigated and discussed at the two collision systems. 
We find, at both RHIC and LHC, the $v_{2}$ mass ordering among $\pi$, $K$, $p$, and $\Omega$ are fairly reproduced within the {\tt VISHNU} hybrid model, 
and more improvements are needed to implement for well describing the $v_{2}$ mass ordering among $p$, $\Lambda$, and $\Xi$.
\end{abstract}

\maketitle

\section{INTRODUCTION}
Ultrarelativistic heavy-ion collisions at the BNL Relativistic Heavy Ion Collider (RHIC) and CERN Large Hadron Collider (LHC) are used to 
produce and study a hot and dense medium consisting of strongly interacting quarks and gluons, namely Quark-Gluon Plasma (QGP), which is expected to exist in the early stage of the universe,
and to understand its properties, such as the equation of state (EoS), transport coefficients.
The hadronic interactions are expected to have less influence on the multi-strange hadrons, such as $\Xi$ and $\Omega$, due to their much smaller hadronic cross sections.
Therefore, final observables of these multi-strange hadrons are more sensitive to the early (partonic) stage of the collision. 
In the past few decades, different aspects of strange and multi-strange hadrons have been investigated theoretically~\cite{Rafelski:1982pu,vanHecke:1998yu,Hamieh:2000tk,Letessier:2000ay,Torrieri:2000xi,Heinz:1998st,Torrieri:2001ue,Huovinen:2001cy,Hiwari:1997vk,Blume:2011sb,Behera:2012eq,Bazavov:2014xya, Zhu:2015sqm}
and experimentally~\cite{Adams:2003fy,Adams:2006ke,Abelev:2007xp,Aggarwal:2010ig,Adams:2005zg,Abelev:2008ae,Adamczyk:2015ukd,Abelev:2013xaa,ABELEV:2013zaa,Abelev:2014pua,Nasim:2015md}.

Anisotropic flow, which is considered as an evidence for the QGP formation, typically displays the collective behavior of the final emitted particles. 
It can be characterized by the coefficients of the Fourier expansion of the final particle azimuthal distribution defined as~\cite{Voloshin:1994mz}:
\begin{equation}
 E\frac{d^{3}N}{d^{3}p}=\frac{1}{2\pi}\frac{d^{2}N}{\pt d\pt dy}(1+2\sum_{n=1}^{\infty} v_{n}\cos[n(\varphi-\Psi_{n})])
 \label{eq:flowVn}
\end{equation}
where $v_{n}$ is the $n^{th}$ order anisotropic flow harmonic with its corresponding reaction plane angle $\Psi_{n}$, and $\varphi$ is the azimuthal angle of the final emitted particles.
Recently, the anisotropic flow and other soft hadron data of all charged and identified hadrons at the RHIC and LHC have been studied by many groups within the framework of 
hydrodynamics~\cite{Zhu:2015sqm,Schenke:2011tv,Qiu:2011hf,Song:2011hk,Song:2011qa,Song:2013qma,Song:2012ua,Petersen:1900zz,Hirano:2005xf,Teaney:2000cw,Bozek:2011wa,Takeuchi:2015s}. 
{\tt VISHNU} is a hybrid model~\cite{Song:2010aq} for single-shot simulations of heavy-ion collisions, which connects the (2+1)-dimensional viscous hydrodynamics with a hadronic afterburner. Employing the {\tt VISHNU} hybrid model, the specific QGP shear viscosity value of $(\eta/s)_{QGP}$ are extracted from the elliptic flow measurements of charged hadrons with {\tt MC-KLN} initial conditions~\cite{Song:2011qa}. With the extracted $(\eta/s)_{QGP}$, the {\tt VISHNU} provides good descriptions of the soft hadron data of $\pi$, $K$, and $p$ at the RHIC and LHC~\cite{Song:2013qma}. Compared with other hadrons, anisotropy flow of (multi-)strange particles are mainly produced in the QGP stage and less contaminated by the subsequent hadronic interactions.
Meanwhile, the $\pt$-spectra and elliptic flow for $\Lambda$, $\Xi$, and $\Omega$ have been measured in the Au+Au collisions at the RHIC~\cite{Aggarwal:2010ig,Adams:2005zg,Abelev:2008ae,Adamczyk:2015ukd} and Pb+Pb collisions at the LHC~\cite{Abelev:2013xaa,ABELEV:2013zaa,Abelev:2014pua}. 
Therefore, it is timely to systematically study these strange and multi-strange hadrons at RHIC and LHC via the {\tt VISHNU} hybrid model.

In this paper, we investigate the $\pt$-spectra and elliptic flow $v_{2}$ for (multi-)strange hadrons in Au+Au collisions at $\sNN=$200 GeV 
and in Pb+Pb collisions at $\sNN=$2.76 TeV within the viscous hydrodynamic hybrid model {\tt VISHNU}. The paper is organized as follows. 
Section~\ref{sec:setup} briefly introduces the {\tt VISHNU} hybrid model and its setup in the calculations. 
Section~\ref{sec:comResults} compares our {\tt VISHNU} results in Au+Au collisions and Pb+Pb collisions with the measurements from the STAR at RHIC and ALICE at LHC, respectively, mainly including $\pt$-spectra and differential elliptic flow for $\Lambda$, $\Xi$, and $\Omega$. 
In Sec.~\ref{sec:massordering}, the mass ordering of elliptic flow among $\pi$, $K$, $p$, $\Lambda$, $\Xi$, and $\Omega$ is studied and discussed at the RHIC and LHC energies. 
Finally, we summarize our works and give a brief outlook for the future in Sec.~\ref{sec:summary}.

\begin{figure*}[t]
  \includegraphics[width=0.8\linewidth,height=9.5cm,clip=]{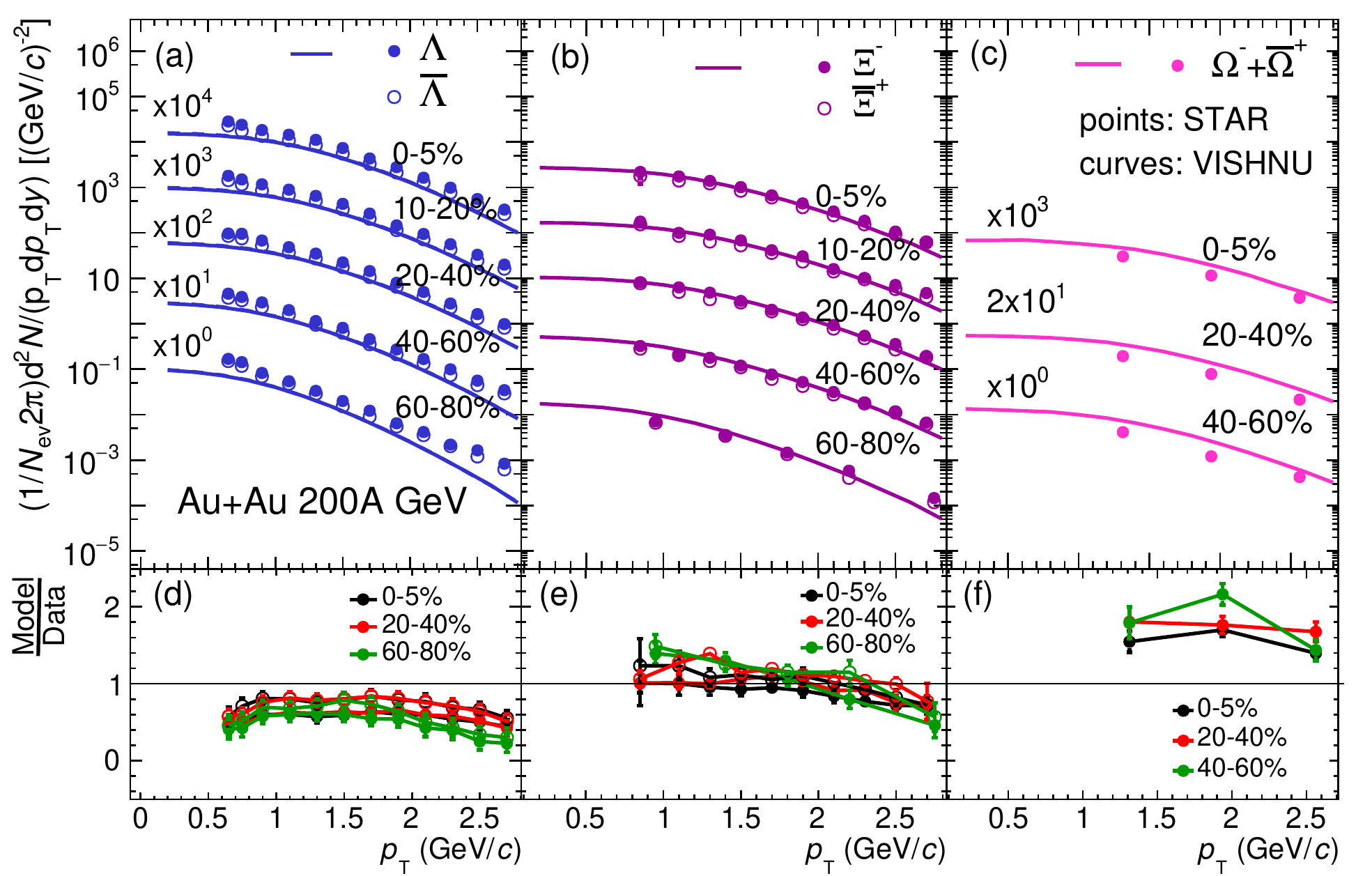}
  \caption{(Color online) Transverse momentum spectra of $\Lambda$($\bar{\Lambda}$), $\Xi^{-}$($\bar{\Xi}^{+}$) and $\Omega^{-}+\bar{\Omega}^{+}$ at various centralities
  in Au+Au collisions at $\sqrt{s_{NN}}$=200 GeV. Experimental data are taken from STAR measurements~\cite{Adams:2006ke}.
  Theoretical curves are calculated by the {\tt VISHNU} hybrid model with the parameters presented in Sec.~\ref{sec:setup}.
  From top to bottom the curves correspond to 0-5\% ($\times10^{4}$), 10-20\% ($\times10^{3}$), 20-40\% ($\times10^{2}$),
  40-60\% ($\times10^{1}$), and 60-80\% ($\times10^{0}$) centrality, respectively, where the factors in parentheses are the multipliers applied to the
  spectra for clear separation. The multiplied factor for spectra of $\Omega$ at 20-40\% is $2\times10^{1}$ instead of $1\times10^{1}$.
  In {\tt VISHNU}, the spectra of particles and corresponding anti-particles are same due to zero net-baryon density used in our calculations. Therefore, only results of particles are shown with solid curves. 
  \label{fig:PtRHIC}}
\end{figure*}
\begin{figure*}[tbph]
 \includegraphics[width=0.8\linewidth, height=9.5cm]{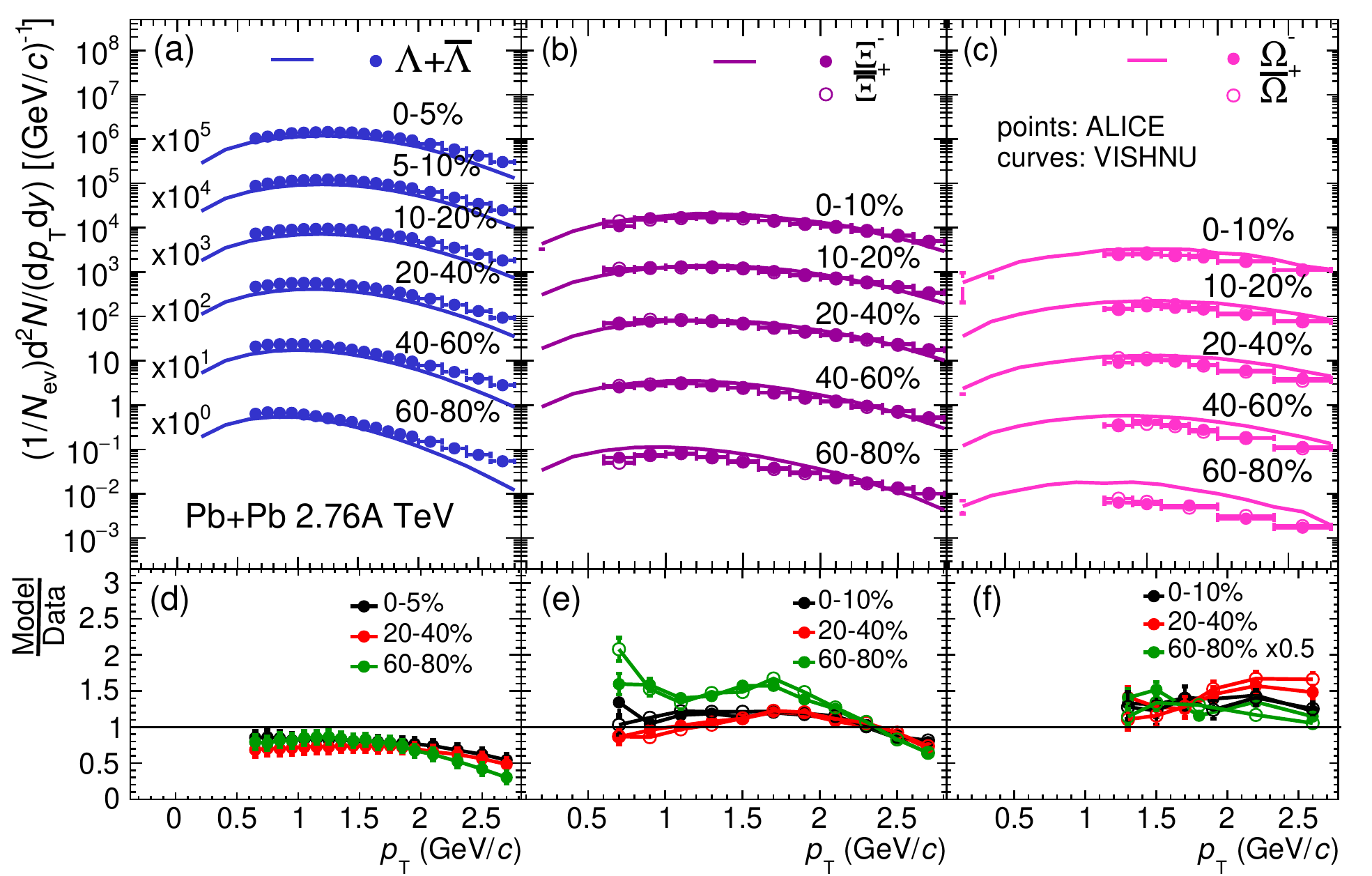}
  \caption{(Color online) Transverse momentum spectra of $\Lambda+\bar{\Lambda}$, $\Xi^{-}$($\bar{\Xi}^{+}$) and $\Omega^{-}$($\bar{\Omega}^{+}$) at various centralities in Pb+Pb collisions at $\sNN=$2.76 TeV. 
  Experimental data are taken from LHC~\cite{Abelev:2013xaa, ABELEV:2013zaa}. Theoretical curves are calculated with the {\tt VISHNU} hybrid model with the parameters presented in Sec.~\ref{sec:setup}. 
  From top to bottom the curves correspond to  0-10\% ($\times10^{4}$), 10-20\% ($\times10^{3}$), 20-40\% ($\times10^{2}$), 40-60\% ($\times10^{1}$) and 60-80\% ($\times10^{0}$) centrality, respectively, 
  where the factors in parentheses are the multipliers applied to the spectra for clear separation. Spectra of $\Lambda$ start from 0-5\% ($\times10^{5}$) and 5-10\% ($\times10^{4}$), instead of 0-10\%.
 In {\tt VISHNU}, the spectra of particles and corresponding anti-particles are same due to zero net-baryon density used in our calculations. Therefore, only results of particles are shown with solid curves. 
 \label{fig:PtLHC}}
\end{figure*}

\section{SETUP OF THE Calculation\label{sec:setup}}
We here give brief descriptions of the inputs and setup of {\tt VISHNU} calculations for the soft data at the RHIC and LHC energies. The {\tt VISHNU}~\cite{Song:2010aq} hybrid model consists of two parts, which are the (2+1)-dimensional ultrarelativistic viscous hydrodynamics {\tt VISH2+1}~\cite{Song:2007fn,Song:2007ux} for the expansion of strongly interacting matter QGP and a microscopic hadronic cascade model ({\tt UrQMD})~\cite{Bass:1998ca,Bleicher:1999xi} for the hadronic evolution.
In the calculations a switching temperature $T_{sw}$ of 165 MeV is set for the transition from the macroscopic to microscopic approaches in {\tt VISHNU}.
This switching temperature value is close to the QCD phase transition temperature~\cite{Aoki:2006br,Aoki:2009sc,Borsanyi:2010bp,Bazavov:2011nk}. 
We input the equation of state ({\tt EOS}) {\tt s95p-PCE}~\cite{Huovinen:2009yb,Shen:2010uy} for the hydrodynamic evolution above the switching temperature $T_{sw}$. The {\tt s95p-PCE}, which accounts for the chemical freeze-out at $T_{chem}$= 165 MeV, was constructed by combing the lattice QCD data at high temperature with a chemically frozen hadron resonance gas at low temperature.

Following Refs.~\cite{Song:2011qa,Song:2013qma}, we input {\tt MC-KLN} initial conditions~\cite{Drescher:2006pi,Drescher:2006ca,Drescher:2007ax} and start the hydrodynamic simulations
at $\tau_0=0.9~{\rm fm}/c$.  For improving computational efficiency, we implement single-shot simulations~\cite{Zhu:2015sqm,Song:2010aq,Song:2011qa,Song:2013qma,Song:2010mg,Zhu:2015dfa} 
with smooth initial entropy density profiles generated by the {\tt MC-KLN} model. The smooth initial entropy densities are obtained by averaging
over a large number of fluctuating entropy density profiles within a specific centrality class. 
The initial density profiles are initialized with the reaction plane method, which was once used in~\cite{Song:2011qa,Song:2013qma,Zhu:2015dfa}. 
Considering the conversion from total initial entropies to final multiplicity of all charged hadrons, we do the centrality selection through the 
distribution of total initial entropies that are obtained from the event-by-event fluctuating profiles. 
Such centrality classification was firstly used by Shen in Ref.~\cite{Shen:2014vra}, which is more close to the experimental one defined
from the measured multiplicity distributions. 
The normalization factors for the initial entropy densities in Au+Au collisions and Pb+Pb collisions are respectively fixed to reproduce the charged hadron multiplicity 
density $dN_{\rm ch}/d\eta$ with $687.4\pm36.6$ at the RHIC~\cite{Adare:2015bua} and $1601\pm60$ at the LHC~\cite{Aamodt:2010cz} at most central collisions. 
The $\lambda$ parameter in the {\tt MC-KLN} model, which quantifies the gluon saturation scale in the initial gluon distributions~\cite{Drescher:2006ca}, 
is tuned to $0.218$ at the RHIC and $0.138$ at the LHC for a better description of the centrality dependent multiplicity density for all charged hadrons.

In the {\tt VISHNU} simulations with {\tt MC-KLN} initial conditions, we set a value of 0.16 for the QGP specific shear viscosity $(\eta/s)_{QGP}$. 
Such combined setting in {\tt VISHNU} calculations once nicely described the elliptic flow of $\pi$, $K$, and $p$ in Au+Au collisions~\cite{Song:2010mg} and Pb+Pb collisions~\cite{Song:2013qma}.
Here, we continue to use it to further study the soft hadron data of strange and multi-strange hadrons at both RHIC and LHC. 
For simplicity the theoretical calculations, we neglect the bulk viscosity, net baryon density, and the heat conductivity in the QGP system evolution.

\section{spectra and elliptic flow \label{sec:comResults}}
In Fig.~\ref{fig:PtRHIC}, we present the transverse momentum spectra of hadrons $\Lambda$, $\Xi$, and $\Omega$ in Au+Au collisions at $\sNN=$200 GeV from the {\tt VISHNU} hybrid model, and compare these results with the STAR measurements. 
We observe that the {\tt VISHNU} generally describes the $p_{\rm T}$-spectra $\Xi$, but slightly overestimates the production of $\Omega$ at all centrality classes. Our {\tt VISHNU} results of $\Lambda$ at $\pt < 2~\gmom$ are about 40\% lower than $\Lambda$ from STAR, and about 20\% lower than $\bar{\Lambda}$ from STAR. This can be understood from following.
In our calculations, the production of $\Lambda$ is obtained from the original values of strong resonance decays from {\tt UrQMD} of {\tt VISHNU}. For the STAR measurements, the $\Lambda$ spectra are corrected for the feed-down of multi-strange baryon weak decays (the feed-down contributions to the $\Xi$ spectra from $\Omega$ decays are negligible)~\cite{Adams:2006ke}. Meanwhile, the STAR $\Lambda$ spectra are not subtracted the contributon from $\Sigma^{0}$ from the channel of $\Sigma^{0}\rightarrow\Lambda+\gamma$\footnote{For $\bar{\Lambda}$, the contribution from $\bar{\Sigma^{0}}$ is via the channel of $\bar{\Sigma^{0}}\rightarrow\bar{\Lambda}+\gamma$.}.
At LHC, it was found that the contribution from $\Sigma^{0}$ for $\Lambda$ is about 30\% in {\tt VISHNU} calculations~\cite{Zhu:2015dfa}.
Furthermore, we notice that STAR measurements of $\Lambda$ and $\Xi^{-}$ are slightly larger than their corresponding anti-particles due to non-zero baryon density at this collision energy.
In our calculations, however, zero net baryon density is used, which leads to the same results between these (multi-)strange hadrons and their anti-particle partners.

\begin{figure*}[tbph]
 \includegraphics[width=0.8\linewidth, height=5.5cm]{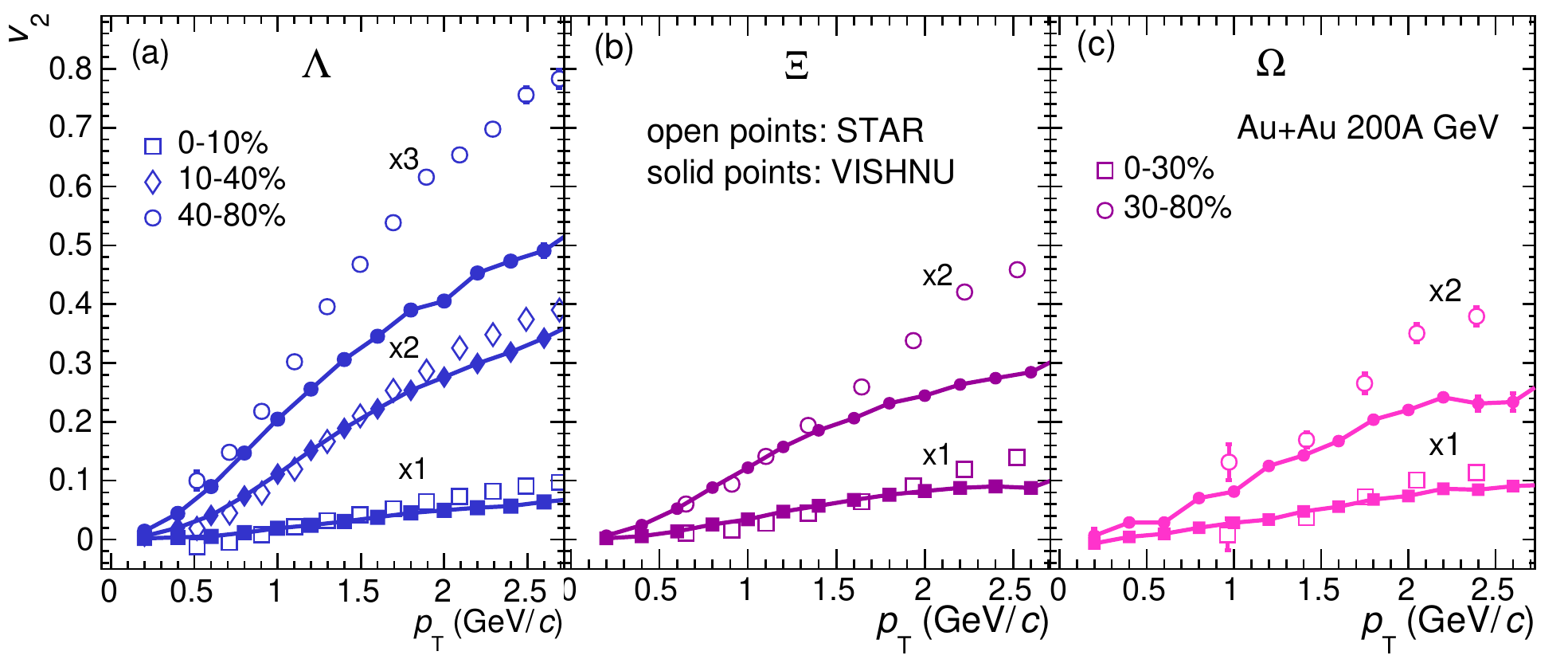}
  \caption{(Color online) Differential elliptic flow of strange hadrons $\Lambda$ (left) at centralities of 0-10\% (squares), 10-40\% (diamonds), and 40-80\%(circles), 
  and multi-strange hadrons $\Xi$ (middle) and $\Omega$ (right) at centralities of 0-30\% (squares), 30-80\% (circles) in Au+Au collisions at $\sNN=$200 GeV.
  Experimental data (open points) are from STAR~\cite{Abelev:2008ae,Adamczyk:2015ukd}. Theoretical results (solid points) are calculated from the {\tt VISHNU} viscous hydrodynamics hybrid model with
  the inputs presented in Sec.~\ref{sec:setup}. 
  \label{fig:V2RHIC}}
\end{figure*}
\begin{figure*}[tbph]
 \includegraphics[width=0.8\linewidth, height=5.5cm]{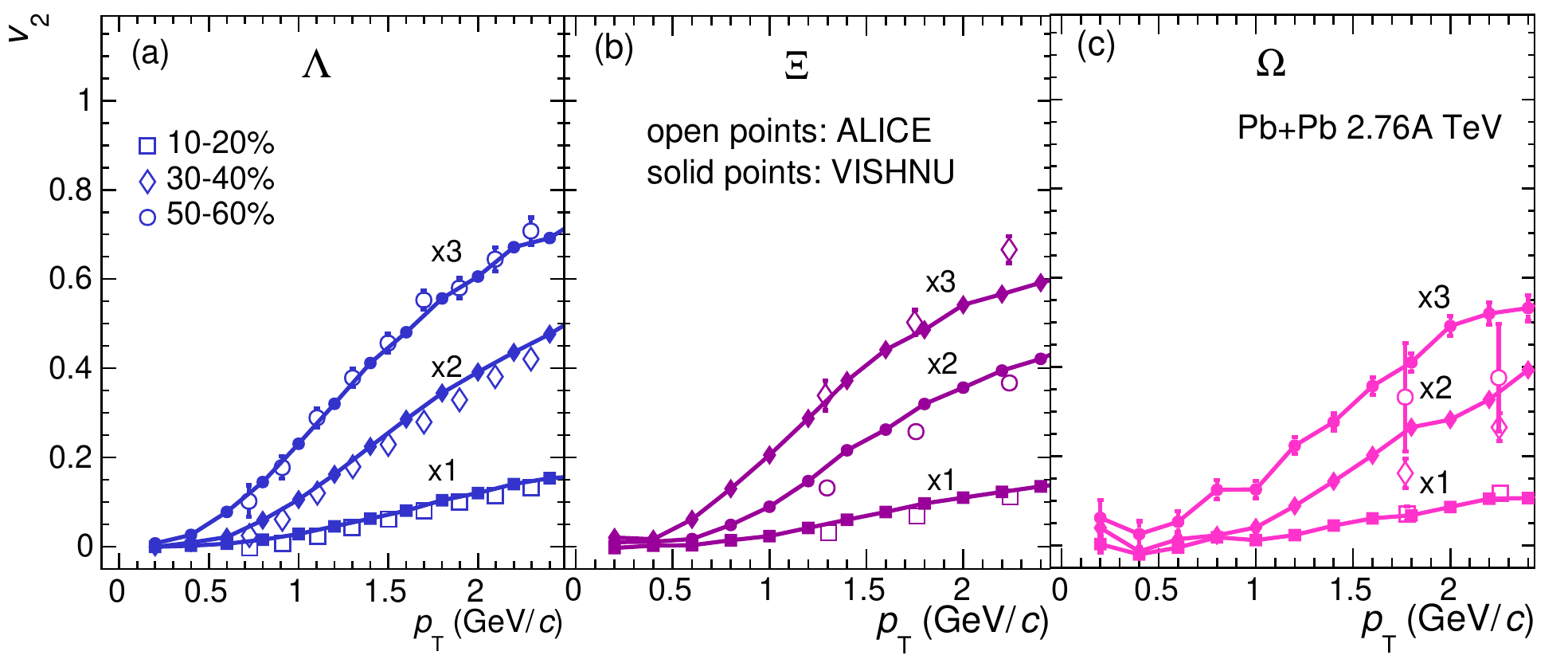}
  \caption{(Color online) Differential elliptic flow of strange hadrons $\Lambda$ (left), multi-strange hadrons $\Xi$ (middle) and $\Omega$ (right) at 10-20\% (squares), 30-40\% (diamonds),
  and 50-60\% (circles) centralities in Pb+Pb collisions at $\sNN=$2.76 TeV. Experimental data (open points) are from the ALICE measurements~\cite{Abelev:2014pua}. 
  Theoretical results (solid points) are calculated from the {\tt VISHNU} hybrid model with the inputs presented in Sec.~\ref{sec:setup}, which were first shown in Ref.~\cite{Zhu:2015dfa}
 \label{fig:V2LHC}}
\end{figure*}

We also calculate the transverse momentum spectra of $\Lambda$, $\Xi$, and $\Omega$ in Pb+Pb collisions at LHC with our {\tt VISHNU} hybrid model. 
The calculations, compared with the measurements from the ALICE Collaboration, are presented in Fig.~\ref{fig:PtLHC}. For the ALICE measurements, 
the production differences between these (multi-)strange hadrons and their anti-particles are very small due to very small net baryon density at the LHC energies. 
It also shows that the {\tt VISHNU} generally describes the $p_{\rm T}$-spectra of hadrons $\Lambda$, $\Xi$, and $\Omega$ at several centrality classes, except for the 60-80\% centrality bin. Here, our theoretical calculations of $\Lambda$ only include strong resonance decays from {\tt UrQMD} of {\tt VISHNU}. 
As a result, they are about 30\% lower than the ALICE measurements including contribution from non-weak decays of $\Sigma^{0}$ and $\Sigma(1385)$ family~\cite{Abelev:2013xaa}.
The $\Omega$ spectra from {\tt VISHNU} are slightly higher than the experimental data at these centrality classes, as similarly observed in the calculations at the RHIC.
Such deviations between theory and experiment are consistent with the model and data differences for the centrality dependent multiplicity shown in~\cite{Zhu:2015dfa}.

From comparisons between our calculations and measurements at the RHIC and LHC, we find that, although the {\tt VISHNU} can not fully reproduce the $\pt$-spectra of these strange and multi-strange hadrons in the production amount, it gives nice descriptions of the slopes (distribution shapes) for the spectra of them at various centralities. This can be found from the ratio between model results and data in Fig.~\ref{fig:PtRHIC} and~\ref{fig:PtLHC}, which are weakly centrality dependent. Together with the early nice descriptions of the $p_{\rm T}$-spectra for $\pi$, $K$, and $p$~\cite{Song:2013qma},
it reveals that during the QGP and hadronic evolution the {\tt VISHNU} hybrid model generates a proper amount of radial flow to push the spectra of various hadrons.

Figure~\ref{fig:V2RHIC} presents the comparisons of differential elliptic flow of $\Lambda$, $\Xi$, and $\Omega$ from the {\tt VISHNU} model with the STAR measurements 
in Au+Au collisions at $\sNN=$200 GeV. The theoretical curves are calculated from {\tt VISHNU} model by using reaction plane initial conditions from the {\tt MC-KLN} model 
and $(\eta/s)_{QGP}=0.16$. The experimental data are from the STAR, which are measured with the event plane method~\cite{Abelev:2008ae,Adamczyk:2015ukd}.
This method covers a fraction of contribution from event-by-event flow fluctuations and non-flow contribution mainly including resonance decays and jets.
Compared with STAR measurements, the elliptic flow from the {\tt VISHNU} generally reproduce the data for $\Lambda$ at 0-10\% and 10-40\%, 
and for $\Xi$ and $\Omega$ at 0-30\% centrality classes.
At 40-80\% Au+Au collisions, the model gives rough descriptions of the data for $\Lambda$ at $\pt < 1.0~\gmom$, and for $\Xi$ and $\Omega$ at $\pt < 1.5~\gmom$, but under-estimates at higher $\pt$ region.
Together with the failed descriptions of elliptic flow for charged hadrons and identified lighter hadrons at 40-80\% collisions in Refs.~\cite{Song:2011hk,Song:2011qa}, 
it reflects that the {\tt VISHNU}, with the {\tt MC-KLN} initial conditions, fails to describe the 40-80\% semi-peripheral collisions. This can be probably interpreted as following. In 40-80\% collisions, the collision lifetimes are shorter, which leave less time to generate the elliptic flow in the fluid dynamic QGP stage, and the highly dissipative effects in hadronic stage cannot compensate for this.

In Fig.~\ref{fig:V2LHC}, we show the differential elliptic flow of $\Lambda$, $\Xi$, and $\Omega$ at 10-20\%, 30-40\%, and 50-60\% Pb+Pb collisions, which were first given in Ref.~\cite{Zhu:2015dfa}.
The presented experimental data are measured by the ALICE Collaboration with the scalar product method~\cite{Abelev:2014pua}. 
The {\tt VISHNU} theoretical results are calculated with the inputs as presented in Sec.~\ref{sec:setup}.
Fig.~\ref{fig:V2LHC} shows that the {\tt VISHNU} fairly predicts the elliptic flow data for $\Lambda$, $\Xi$, and $\Omega$ at chosen three centrality classes at $\pt < 2~\gmom$ within the statistical error bars.
At $\pt > 2~\gmom$, the descriptions of the elliptic flow for $\Xi$ at 50-60\% and for $\Omega$ at 30-40\% and 50-60\% become worse. 
Together with the worse descriptions of elliptic flow data at high-$\pt$ at the RHIC, as shown in Fig.~\ref{fig:V2RHIC}, we consider that 
the viscous corrections probably become too large at high-$\pt$ region, in which the hydrodynamic description in the {\tt VISHNU} lost its predictive power.

\section{mass ordering of elliptic flow\label{sec:massordering}}
It is widely accepted that the characteristic mass ordering of differential elliptic flow among various identified hadrons at low-$\pt$ 
reflects the interplay between radial and elliptic flow, providing more insights into the properties of the QGP fireball. 
The radial flow creates a depletion in the particle $p_{\rm T}$-spectrum at low values, which increases with increasing particle mass. 
This leads to heavier particles having a smaller $v_{2}$ compared to lighter ones at a given value of $p_{\rm T}$, giving a mass ordering of the $p_{\rm T}$ dependent elliptic flow below $1.5-2~\gmom$. 
Such $v_2$ mass ordering has been discovered in the experiments at both the RHIC and LHC~\cite{Abelev:2008ae,Abelev:2014pua,Adams:2004bi,Issah:2006qn,Snellings:2014vqa}, 
which has also been studied within the framework of hydrodynamics~\cite{Huovinen:2001cy,Zhu:2015sqm,Hirano:2007ei,Bozek:2011fm,Song:2013tpa} and blastwave model~\cite{Huovinen:2001cy,Adler:2001nb}.
 
\begin{figure}[tbph]
  \includegraphics[width=1\linewidth, height=9cm]{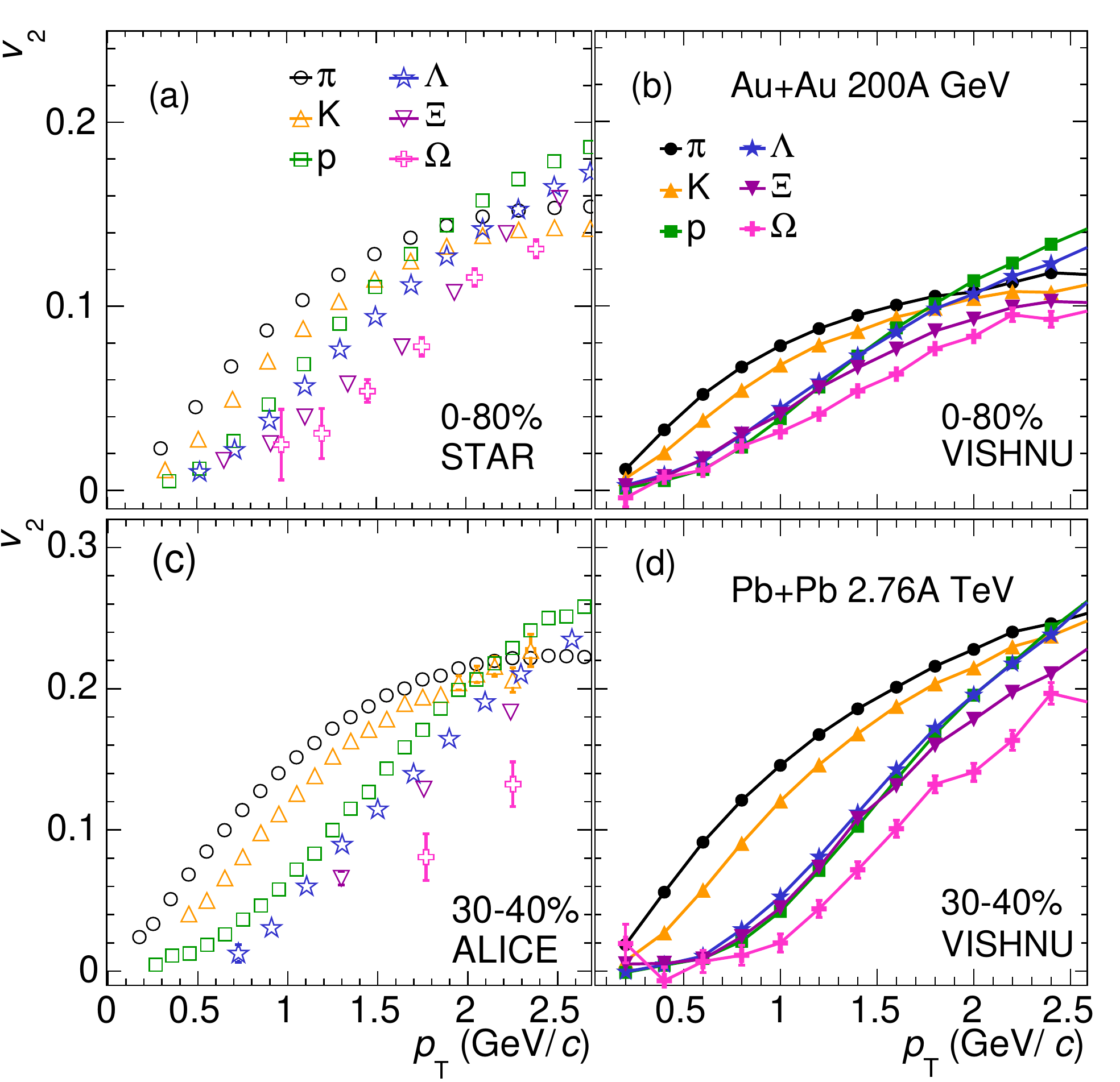}
   \caption{(Color online) Differential elliptic flow of $\pi$, $K$, $p$, $\Lambda$, $\Xi$, and $\Omega$ at centrality 0-80\% 
   in Au+Au collisions at $\sNN=$200 GeV (top two panels, left: data, right: {\tt VISHNU} calculations), and at centrality 30-40\% in Pb+Pb collisions at $\sNN=$2.76 TeV (bottom two panels, left: data, right: {\tt VISHNU} calculations).
   The measurements of elliptic flow of various hadron species are taken from STAR~\cite{Adams:2006ke,Adamczyk:2015ukd} and ALICE~\cite{Abelev:2014pua}.
   \label{fig:V2mass}}
\end{figure}
Here we investigate the mass ordering of the elliptic flow among various identified hadrons $\pi$, $K$, $p$, $\Lambda$, $\Xi$, and $\Omega$ in Au+Au collisions and Pb+Pb collisions.
For clear presentations, the experimental data and our {\tt VISHNU} results are plotted in separate panels at 0-80\% at the RHIC and 30-40\% at the LHC.
Together with the calculations of elliptic flow of identified hadrons in~\cite{Song:2011qa, Song:2013qma} and in this paper, 
we find that the {\tt VISHNU} generally describes the $v_2(p_T)$ for different identified hadrons at several centrality classes.
However, our theoretical results presented in Fig.~\ref{fig:V2mass}, compared with the measurements at RHIC and LHC, show that the {\tt VISHNU} fairly describes the mass ordering among
$\pi$, $K$, $p$, and $\Omega$. But it is hard to see mass ordering among $p$, $\Lambda$ and $\Xi$ clearly due to their elliptic flow being almost identical at low $\pt$ region. 
This under-prediction for proton leads to an inverse $v_2$ mass ordering between $p$ and $\Lambda$. 
Effects of hadronic rescattering on elliptic flow are seen to be particle specific, depending on their scattering cross sections that couple them to the medium~\cite{Hirano:2007ei}.
Compared with non-strange hadrons, the multi-strange hadrons are less affected on their differential elliptic flow due to their smaller scattering cross sections. 
Therefore, re-evaluating the hadronic cross sections in the {\tt UrQMD} is helpful to improve the description of elliptic flow of various hadron species.
Meanwhile, an initial flow could enhance the radial flow in the hadronic stage, which is also expected to improve the description of mass ordering 
within the framework of the hybrid model.

\section{Summary and outlook\label{sec:summary}}
In summary, we studied the $\pt$-spectra and elliptic flow of strange and multi-strange hadrons in Au+Au and Pb+Pb collisions within the {\tt VISHNU} hybrid model.
At both collision systems, we found that, with {\tt MC-KLN} initial conditions, $\eta/s =0.16$ and other inputs, {\tt VISHNU} generally describes the $\pt$-spectra of 
strange hadron $\Lambda$ and multi-strange hadrons $\Xi$ at some centrality classes, but slightly over-estimates for $\Omega$ at chosen centrality classes.
In spite of the normalization issues, the {\tt VISHNU} well produces the spectra slopes of these three hadrons at chosen centralities.
By comparing the elliptic flow of $\Lambda$, $\Xi$, and $\Omega$ in Au+Au collisions and Pb+Pb collisions from {\tt VISHNU} model with the STAR and ALICE measurements,  
we found that the {\tt VISHNU} generally describes the elliptic flow except at 40-80\% semi-peripheral collisions. 
The failed descriptions at 40-80\% collisions is probably due to shorter lifetimes of these collisions, which leave less time to generate the elliptic flow in the fluid dynamic QGP stage.

We also compared the mass ordering of $v_{2}$ among hadrons $\pi$, $K$, $p$, $\Lambda$, $\Xi$, and $\Omega$ from {\tt VISHNU} calculations with the STAR and ALICE measurements. 
The comparisons showed that the elliptic flow mass ordering among various hadron species is not fully described at both RHIC and LHC. 
The {\tt VISHNU} fairly describes the mass ordering of $v_{2}$ among $\pi$, $K$, $p$, and $\Omega$, but fails to reproduce the mass ordering among $p$, $\Lambda$, and $\Xi$
due to slight under-predictions of the elliptic flow of protons. The effects from the initial flow and/or improved {\tt UrQMD} hadronic cross-sections
may solve this issue within the framework of {\tt VISHNU}, which should be investigated in the near future. 

\section*{Conflict of Interests}
The author declares that there is no conflict of interests regarding the publication of this paper.

\acknowledgments
\vspace*{-2mm}
The author gratefully thanks Huichao Song and Hao-jie Xu for fruitful discussions and critical reading of the draft.
This work was supported in part by the NSFC and the MOST under Grants No. 11435001 and No. 2015CB856900, 
and the China Postdoctoral Science Foundation under Grant No. 2015M570878.  
The author especially acknowledges extensive computing resources provided by Tianhe-1A from the National Supercomputing Center in Tianjin, China.

\end{document}